# RELATIVISTIC ELECTRON BEAM SLICING BY WAKEFILED IN PLASMAS

S. V. Bulanov[1,†], T. Tajima[1], G. Mourou[2]

[1]*Kansai Photon Science Institute, JAEA, Kizu-cho, Kyoto-fu 619-0215, Japan*

[2]*LOA ENSTA/Ecole Polytechnique, F-91761 Palaiseau CEDEX, France*

*ABSTRACT*: A method of slicing of high-energy electron beams following their interaction with the transverse component of the wakefield left in a plasma behind a high intensity ultra short laser pulse is proposed. The transverse component of the wakefield focuses a portion of the electron bunch, which experiences betatron oscillations. The length of the focused part of the electron bunch can be made substantially less than the wakefield wavelength.

1. **Introduction**

An important scientific problem in future accelerators is the generation of extremely short relativistic electron bunches. Electron bunches of a femtosecond duration can be employed for a myriad of applications, including the femtosecond x-ray pulse generation, in particular, for the purposes to investigate the matter dynamics in the interatomic time scale and the rapid biological dynamics.

The utilization of intense laser to accelerate electrons in a compact fashion through laser wakefield is well known [1]. Here we introduce a method to efficaciously utilize the wakefield for the purpose of focusing the accelerated beam. As well known, the laser radiation produces the ponderomotive force that acts on charged particles and allows the controlled manipulation of electrons. The ponderomotive acceleration of electrons by strong electromagnetic force in vacuum

---

[†] Also at A. M. Prokhorov Institute of General Physics of Russian Academy of Sciences, 119991 Moscow, Russia

has been proposed in Ref. [2] and it has been discussed actively since then [3-6]. The ponderomotive force is capable of trapping charged particles, as it has been observed in the experiment presented in Ref. [7]. A combination of the ponderomotive electron trapping in the transverse direction and their acceleration in the longitudinal direction has been discussed in Refs. [5,6], where the relativistic electron bunch slicing by the ponderomotive force in both the longitudinal and transverse directions has been described in detail. We notice here the relativistic electron bunching produced as a result of its interaction with the laser pulse during the electron propagation inside the wakefield wiggler. An ultrashort laser pulse in this case modulates the electron energy and then energy modulated electrons are spatially separated by the magnetic field. This technique has been used in Ref. [8] in order to generate femtosecond x-ray pulses. The relativistic electron bunching may also occur as a result of the radiation instability development as it has been shown in Ref. [9].

Due to the very high electric field it can sustain, plasma may be used as a medium for the relativistic electron beam slicing [10] in the longitudinal direction. The plasma lens has also been proposed as a final focusing element to improve the luminosity of future high energy electron-positron colliders [11]. When the laser pulse propagates in an underdense plasma, its ponderomotive pressure forms regions with strong electric charge separation, which propagates with the speed of the laser pulse. Charged particle acceleration by these wake plasma waves left in a plasma behind a high intensity ultra short laser pulse has been attracting a great attention for a long time because wake waves can have electric field that is orders of magnitude higher than those found in conventional accelerators [1]. The longitudinal component of the electric field contributes to the acceleration of charged particles and the transverse component affects the particle motion in the direction perpendicular to the laser pulse propagation (see Figs. 2 and 6 in Ref. [12]).

In order to examine the detailed structure of the electric field inside the first period of the wake wave, we carried out 2D PIC simulations with the code REMP based on the "density decomposition" scheme [13]. In the simulations, the laser pulse linearly polarizes along the *z* axis; it propagates along the *x* axis. Its dimensionless amplitude is *a* = 5, corresponding to the peak intensity *I* = 5.36×10$^{18}$ (0.8 μm/λ)2 W/cm$^2$. Here, *x* and *y* are normalized to the laser pulse wavelength λ = 2π*c*/ω; the time unite is equal to the laser radiation period, *T* = 2π/ω; and the plasma density is normalized to the critical density, $n_{cr}$ = *m*eω$^2$/4π*e*$^2$. The laser pulse has a size of 10×60. The laser pulse interacts with a 225×150 plasma slab with the density *n* =1×10$^{-4}$ $n_{cr}$. In Fig. 1, we present the results of simulations. Here, we show the electron density distribution (a), the x (b) and y (c) components of the electric field, in the (*x*, *y*) plane for *t* = 200. We see that the x and y components of the electric field in the vicinity of the laser pulse axis are linear functions of the coordinates x and y, respectively.

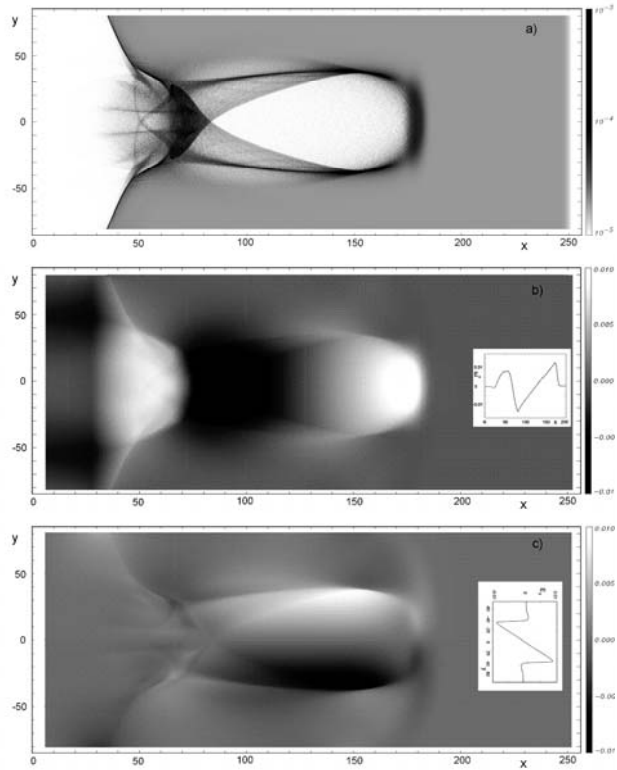

**Figure 1.** Wake wave structure in the first period of the wave behind the laser pulse for t=200. (a) The electron density distribution in the (x,y) plane. (b) The x component of wakefield in the (x,y) plane; the slice of the x component of wakefield at y=0 is shown. (c) The y component of wakefield in the (x,y) plane; the slice of the y component of wakefield at x=150 is shown.

Within a substantially long portion of the wakefield the transverse electric field focuses electrons toward the axis. As a result, an unmatched beam oscillates with the betatron frequency, periodically focusing and defocusing. In Ref. [14] observed were radial jets of low energy electrons, which were likely caused by the transverse ejection of electrons due to the radial structure of the wakefield and space charge deflection of electrons, as they exit the laser focus. Multiple betatron oscillations of the ultra-relativistic electron beam caused by the transverse component of the beam excited wake field have been studied experimentally in Ref. [15]. The length of the focused part of the electron beam is approximately equal to the wavelength of the wake wave. If the plasma slab length is chosen to be slightly less than the betatron oscillation length, the electron bunch is focused in a vacuum region at the rear side of the slab. By using a narrow collimator, we can extract a focused portion of the electron beam. This opens a way for slicing ultrarelativistic electron beams into sub-picosecond beamlets.

The proposed scheme is illustrated in Fig. 2. In this simplified picture the electron beam slicer consists of a finite length underdense plasma slab and a narrow collimator. At the first stage shown in Fig. 2 a, a relativistically intense laser pulse generates the wake wave. Inside the wake wave there is a transverse electric field which focuses the co-propagating electron beam. Then, at the second stage explained in Fig. 2 b, the electron beam leaves the plasma slab with the negative radial velocity and continues to focus. At the position of the focus we put the collimator with the pin-hole diameter of the order of the focused beam waist. After the beam has passed through the collimator it appears at the rear side in the form of a beamlet with the length less than the wake wave wavelength (see Fig. 2 c).

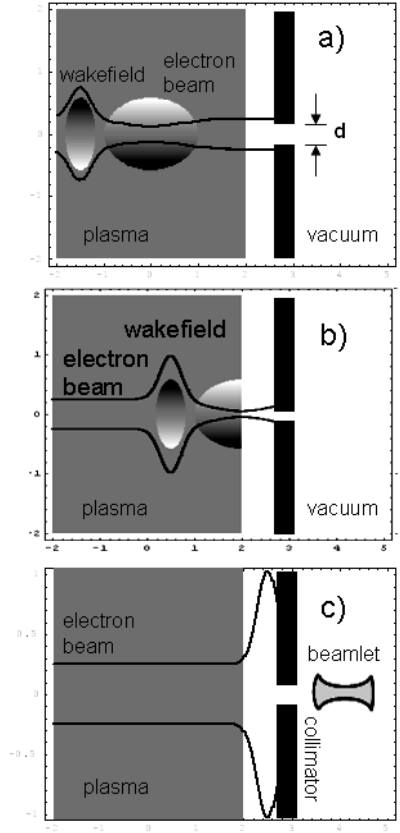

**Figure 2.** The slicer scheme. a) The electron beam is modulated by the wakefield inside the plasma slab; its parts are focused and defocused by the transverse component of the wake electric field; b) The focused electron beam enters the vacuum region; c) The collimator cuts off the short and narrow electron beamlet.

## 2. Mechanism of Relativistic Electron Beam Slicing

Plasma waves driven in plasmas by ultra short laser pulses have typical accelerating gradients of the order of or greater than $eE_{\|} = 100\, GeV/m$. For the wake field wavelength equal to $\lambda_{wf} \approx 30\,\mu m$ it corresponds to the wake electrostatic potential $\varphi \approx E_{\|}\lambda_{wf}/2\pi = 0.5\, MV$. Due to a difference between the phase velocity of the wake field, $v_{ph}$, and the velocity of the relativistic charged particle, $v \approx c$, the particle energy gain is $\Delta \mathcal{E} = e\varphi/(1 - v_{ph}/c) \approx 2e\varphi \gamma_{ph}^2 = eE_{\|}\lambda_{wf}\gamma_{ph}^2/\pi$, where $\gamma_{ph} = (1 - v_{ph}^2/c^2)^{-1/2}$. In the case when the transverse size of the sufficiently intense laser pulse, $r_0$, is less than the plasma wavelength, $\lambda_{pe} \approx 2\pi c/\omega_{pe}$ with $\omega_{pe} = (4\pi n_0 e^2/m_e)$, the electrostatic potential is of the order of $\varphi \approx \pi n_0 e r_0^2$. Near the axis the radial electric field is a linear function of the radius: $E_\perp = 2\pi n_0 e r$.

*2.1. Description of Betatron Oscillations with the Envelope Equation*

If the electron energy gain $\Delta\mathcal{E}$ is much less than the electron initial energy $\mathcal{E} = m_e c^2 \gamma_b$ the transverse dynamics of the electron beam with the normalized transverse emittance $\varepsilon_N$ is described by the beam-envelope equation [16] for the bunch radius $\sigma_r$:

$$\frac{d^2\sigma_r(x)}{dx^2} + Q_b^2 \sigma_r(x) - \frac{\varepsilon_N^2}{\gamma_b^2 \sigma_r^3(x)} = 0, \tag{1}$$

where we have neglected effects of space charge and beam generated magnetic field. Here $Q_b = \omega_b/c$ with $\omega_b = \omega_{pe}/(2\gamma_b)^{1/2}$ being the betatron oscillation frequency. In order to solve Eq. (1), we introduce dimensionless variables, $\xi = Q_b x$ and $\sigma = (\gamma_b Q_b/\varepsilon_N)^{1/2} \sigma_r$, and rewrite Eq. (1) as

$$\sigma'' + \sigma - \frac{1}{\sigma^3} = 0, \tag{2}$$

where a prime denotes differentiation with respect to $\xi$. This equation has an integral

$$(\sigma')^2 + \sigma^2 + \frac{1}{\sigma^2} = q_p. \tag{3}$$

Here $q_p = (\sigma'_0)^2 + \sigma_0^2 + \sigma_0^{-2}$. It is easy to see that $q_p \geq 2$. From Eq. (3) we find

$$\sigma = \left\{ \frac{1}{2} \left[ q_p \pm \left(q_p^2 - 4\right)^{1/2} \sin\left( 2\xi + \text{ArcTan} \frac{q_p - 2\sigma_0^2}{2|\sigma'_0 \sigma_0|} \right) \right] \right\}^{1/2}, \tag{4}$$

where a sign in front of the second term in the square brackets on the right had side of the equation is determined by the sign of $\sigma'_0$.

We see that for unmatched beam (the matched beam with $q_p = 2$, $\sigma'_0 = 0$, propagates with the constant waist, $\sigma = 1$) its waist changes between $\sigma_{min} = \left\{\left[q_p - (q_p^2 - 4)^{1/2}\right]/2\right\}^{1/2}$ and $\sigma_{min} = \left\{\left[q_p + (q_p^2 - 4)^{1/2}\right]/2\right\}^{1/2}$. For $q_p = 2 + \delta$ with $\delta \ll 1$ we have $\sigma_{min} \approx \left[1 - (\delta/2)^{1/2}\right]^{1/2}$ and $\sigma_{min} \approx \left[1 + (\delta/2)^{1/2}\right]^{1/2}$. In the limit $q_p \gg 2$ we obtain $\sigma_{min} \approx (1/q_p)^{1/2}$ and $\sigma_{max} \approx q_p^{1/2}$.

## 2.2. Beam-Envelope Dynamics in the Defocusing Phase of Wakefield

In the defocusing phase of the wakefield, where the transverse component of the electric field is negative, Eqs. (2) and (3) change to

$$\sigma'' - \sigma - \frac{1}{\sigma^3} = 0, \tag{5}$$

and

$$(\sigma')^2 - \sigma^2 + \frac{1}{\sigma^2} = q_d, \tag{6}$$

where $q_d = (\sigma'_0)^2 - \sigma_0^2 + \sigma_0^{-2}$. From Eq. (6) we obtain

$$\sigma = \left\{\frac{1}{2}\left[(q_d + 2\sigma_0^2)\cosh 2\xi \pm 2|\sigma_0'\sigma_0|\sinh 2\xi - q_d\right]\right\}^{1/2}, \tag{7}$$

i.e. at $\xi \to \infty$ we have $\sigma \approx \left[(q_d + 2\sigma_0^2)^{1/2}/2\right]e^\xi$. Electrons from the defocusing phase of the wakefield are dispersed wide.

## 2.3. Beam Focusing in Vacuum Region

The beam waist after electrons leave the plasma slab at $\xi = L_p$ and enter the vacuum region is described by equation

$$\sigma'' - \frac{1}{\sigma^3} = 0, \tag{8}$$

whose solution reads

$$\sigma = \left\{ \frac{1}{q_v} \left[ 1 + \left( |\sigma_L' \sigma_L| \pm q_v \xi \mp q_v L_p \right)^2 \right] \right\}^{1/2}. \tag{8}$$

Here $q_v = (\sigma'_L)^2 + \sigma_L^{-2}$, $\sigma_L' = \sigma'|_{\xi=L_p}$, and $\sigma_L = \sigma|_{\xi=L_p}$. For negative $\sigma_L'$ the beam waist decreases to its minimal value

$$\sigma_{min} = \left( \frac{1}{q_v} \right)^{1/2} = \frac{\sigma_L}{\left[ 1 + (\sigma'_L \sigma_L)^2 \right]^{1/2}} \tag{9}$$

at the distance $\xi_f = |\sigma_L' \sigma_L| / q_v$ from the plasma slab - vacuum interface. After the focus region the electron bunch expands when its waist increases with the distance according to the asymptotic dependence given by expression: $\sigma \approx q_v^{1/2} \xi$. Related discussion may be found in Ref. [17]. Typical electron beam waist dependence on the coordinate $\xi$ is plotted in Fig. 3.

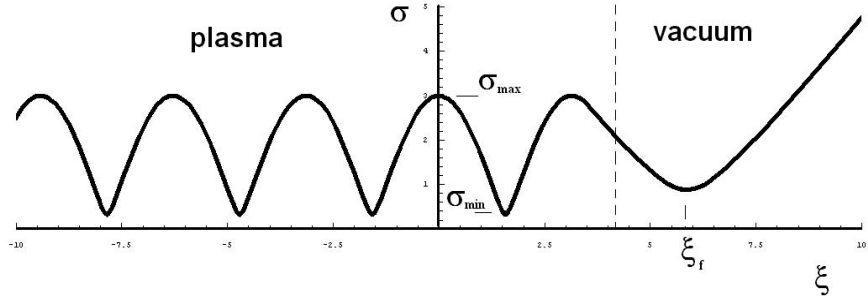

**Figure 3.** Betatron oscillations inside the plasma slab and the electron beam focusing and defocusing in the vacuum region for $\sigma_0 = 3, \sigma_0' = 0$ and $L_p = 3.5125$.

## 2.4. Cutting off the Short Beamlet with the Collimator

From Eq. (4) we can find that a good focusing occurs when $\sigma_L' \approx \sigma_L \approx \sigma_0$. In this case $\sigma_{min} \approx 1/\sigma_0$ and $\xi_f \approx 1$. In dimensional unites we have $\sigma_{r,min} = \varepsilon_N / \gamma_b Q_b \sigma_{r,0}$ and $x_f \approx 1/Q_b$. Using the collimator with a hole diameter, $d \ll x_f$, we can cut the electron bunch on the beamlets of the length $l_{bt} = x_f d / \sigma_{r,0}$. The shortest beamlet is produced when $d = 2\sigma_{r,min}$; its length in this case is equal to $l_{bt,min} = 2\varepsilon_N / \gamma_b Q_b^2 \approx 4\varepsilon_N c^2 / \omega_{pe}^2$.

Since the focusing length $x_f \approx c(2\gamma_b)^{1/2}/\omega_{pe}$ and the focus waist $\sigma_{r,min} = 2^{1/2}\varepsilon_N c / \gamma_b^{1/2} \omega_{pe} \sigma_{r,0}$ depend on the particle energy by changing the collimator position and its hole diameter $d$ we can extract a narrow energy beamlet from the electron beam with a wide energy distribution.

We assume the plasma density to be $10^{18} cm^{-3}$, which corresponds to $Q_b = 141 cm^{-1}$ for $0.8\mu m$ laser, $\gamma_b = 100$ and the transverse emittance $\varepsilon = 3.5 \times 10^{-2}$ mm-mrad, and the electron beam initial radius $\sigma_{r,0}$ equal to $20\mu m$. From above written expressions we find the focus waist $\sigma_{r,min} = 0.7\mu m$ and the shortest beamlet length $l_{bt,min} = 5\mu m$.

## 3. Discussion and Conclusions

We here presented an approach for slicing of high-energy electron beams due to focusing of the electron by the transverse component of the wakefield generated in a plasma slab by a high intensity laser. Then, the electron beam leaves the plasma slab with the negative radial velocity and continues to focus. At the position of the focus the collimator with the pin-hole diameter of the order of the focused beam waist, after the beam has passed through the collimator it appears at the rear side in the form of a beamlet with the length less than the wake wave wavelength. We note that the present method is efficient because the focusing transverse field propagates with the electron beam, thereby elongating the interaction time by a factor of $\gamma_{ph}$. The current method may constitute of novel ways in what we call relativistic engineering. The intensity of the wakefield focusing field amounts to 100 GeV/m, allowing the effectiveness in extreme high energy applications as well as efficient compact focusing.

As compared to vacuum focusing method [5], where focusing force arises due to the transverse component of the ponderomotive force or/and the laser transverse fields in a capillary tube, this focal force is subject to the breakdown of the surface material of the capillary and surely much less than GeV/m, order of magnitude smaller than what we consider here.

We also note an analogy between the present method for the electron beam slicing and the approach for manipulating the laser accelerated ion beam quality by using the electrostatic focusing system. Within the framework of the ion focusing technique [18] a thin hollow cylindrical shell is irradiated by a femtosecond high-power laser pulse when the ion bunch flies through it. In their method the heated plasma in the cylinder ablates inwards. Due to the electric charge separation at the plasma front the radial electric field is generated, which affects on the transverse ion motion similar to the effect produced by the transverse component of the wake field on fast electrons. As experimentally demonstrated in [18], this technique allows one to simultaneously focus the proton beam and to cut it onto quasi-monoenergetic beamlets.

Extremely short bunches thus created through the presented method may be beneficial to many applications, including femtosecond and attosecond X-ray generations, fast time-resolved radiolysis, and high energy accelerators.

**Acknowledgments**

This work was partly supported by the Ministry of Education, Culture, Sports, Science and Technology of Japan, Grant-Aid for Specially Promoted Research No. 15002013.